\documentclass[prb,aps,twocolumn,superscriptaddress,nofootinbib]{revtex4-2}

\usepackage{amsmath}
\usepackage{amssymb}
\usepackage{mathrsfs}
\usepackage[bbgreekl]{mathbbol}
\usepackage{mathtools}
\usepackage{physics}
\usepackage[american]{babel}

\usepackage{multirow}
\usepackage{hhline}

\usepackage[x11names]{xcolor}
\usepackage[colorlinks=true, citecolor=blue!90!black, linkcolor=blue!90!black, linktocpage=true, urlcolor=red!70!black]{hyperref}

\usepackage{tikz}
\usetikzlibrary{decorations.pathreplacing}

\definecolor{MajBlue}{HTML}{377eb8}
\definecolor{MajRed}{HTML}{e41a1c}

\renewcommand{\eqref}[1]{Eq.~(\ref{#1})}

\makeatletter
\newcommand{\HideAppendixSubsectionsFromToc}{%
  \let\l@subsection\@gobbletwo
  \let\l@subsubsection\@gobbletwo
}
\makeatother



\begin{document}

\hfill MIT-CTP/6134

\title{An Exact QFT/Lattice Map for Fermions}
\author{Shu-Heng Shao}
\affiliation{
Center for Theoretical Physics -- a Leinweber Institute,
Massachusetts Institute of Technology,
Cambridge, MA 02139, USA
}

\author{Nikita Sopenko}
\affiliation{
Center of Mathematical Sciences and Applications,
Harvard University,
Cambridge, MA 02138, USA
}

\begin{abstract}
We construct an exact map between continuum fermion fields and lattice fields in 1+1 dimensions. This map rewrites a continuum Hamiltonian as a lattice Hamiltonian with nearest-neighbor couplings and an infinite-dimensional local Hilbert space. The resulting lattice model reproduces exactly the energy spectrum, global operator algebra, chiral  symmetries, and anomalies of the continuum QFT on any finite lattice. 
Unlike domain wall fermions, truncating the local Hilbert space introduces no extra massless modes, and the entanglement entropy of an interval converges as the truncation is removed. 
The lattice local algebras are of type I and lie between the type III continuum algebras of nested spatial regions, in accordance with the split property.   
We demonstrate this construction explicitly for $N$ Majorana fermions, realizing the  $\mathrm{O}(N)_\mathrm{L}\times\mathrm{O}(N)_\mathrm{R}$ lattice chiral symmetry. 
We also construct an exact lattice model for the Schwinger model and numerically reproduce the mass gap $e/\sqrt{\pi}$. 

\end{abstract}

\maketitle
\tableofcontents

\section{Introduction}

Given a quantum field theory, such as the Standard Model, can we formulate it on a lattice? 
This is a longstanding and fundamental question, motivated not only by its conceptual significance but also by the practical importance of lattice formulations for numerical simulations. 
The question becomes even more challenging when one further requires the lattice model to preserve the chiral global symmetries of the QFT. Realizing these symmetries on the lattice is essential for constructing chiral gauge theories by subsequently gauging anomaly-free subgroups. 
The Nielsen--Ninomiya  theorem \cite{Nielsen:1980rz,Nielsen:1981xu,Nielsen:1981hk,Friedan:1982nk,Shamir:1993bi} states that it is impossible to realize the chiral U(1) global symmetry of a Weyl fermion in 3+1d using lattice free fermions. 
More recently, such no-go theorems were generalized in \cite{Fidkowski:2023sif,Kapustin:2024rrm,Liu:2026atf} for anomalous continuous global symmetries in general interacting Hamiltonian lattice systems.\footnote{The original Nielsen--Ninomiya theorem was formulated in Hamiltonian lattice models where space is discrete and time is continuous. See \cite{Karsten:1980wd,Karsten:1981gd,Baig:2026zgb} for the Euclidean counterparts  of these no-go theorems for spacetime lattice models, which are formulated under different settings.} 

Lying at the heart of these no-go theorems for Hamiltonian lattice models is the assumption that the local Hilbert space is finite-dimensional, as is the case for lattice fermions and qubits. However, all bets are off when the local Hilbert space is infinite-dimensional.\footnote{Important exceptions are QFTs with pure gravitational anomalies, such as a single right-moving chiral fermion in 1+1 dimensions, which is believed to be impossible to realize on a lattice even with an infinite-dimensional local Hilbert space. See \cite{Jones:2026gmd} for a recent discussion and, e.g., \cite{Hellerman:2021fla}, for earlier discussions. 
App.\ \ref{app:chiral} presents a formal lattice model for a chiral fermion, but it is not a genuine 1+1d model and is qualitatively similar to the domain wall fermion. 
In contrast, a Weyl fermion in 3+1 dimensions, without imposing its U(1) global symmetry, has no gravitational anomaly. Indeed, it can be realized on a lattice \cite{Nielsen:1981xu,Gioia:2025bhl}.} \textit{Infinite is different.}
Indeed, Hamiltonian lattice models with an infinite-dimensional bosonic Hilbert space at each site, including the modified Villain models \cite{Cheng:2022sgb,Fazza:2022fss,Seifnashri:2026ema,Lu:2026jnq,Lu:2026itw,2026arXiv260802722D} and rotor models \cite{Thorngren:2026ydw,Fidkowski:2025rsq,2026arXiv260802722D}, have been constructed in 1+1 and 3+1 dimensions, realizing continuous abelian chiral global symmetries and their 't Hooft anomalies. These Hamiltonian constructions build on earlier studies of Euclidean modified Villain lattice models \cite{Gross:1990ub,Sulejmanpasic:2019ytl,Gorantla:2021svj,Berkowitz:2023pnz,Baig:2026zgb} (see also \cite{DeMarco:2017gcb,Demarco:2018fuw,Chen:2019mjw,DeMarco:2023hoh}).
However, these models do not realize non-abelian chiral global symmetries, leaving open the challenge of constructing more faithful lattice realizations of chiral QFTs.

How well can we approximate a QFT using lattice models? 
In this paper, we construct an exact map between continuum quantum fields and lattice fields, which rewrites a continuum Hamiltonian as a lattice one. The resulting lattice Hamiltonian has nearest-neighbor couplings and an infinite-dimensional local Hilbert space. It shares the same  spectrum, global symmetries, and anomalies with the continuum QFT even at finite system size. 
In particular, although not manifest, the lattice model has the exact Poincar\'e spacetime symmetry. 
Such exact QFT/Lattice maps have been shown to exist for 1+1d free fermions and diagonal rational conformal field theories in \cite{Jones:2026gmd}, though the construction there is rather implicit. In this paper, we provide explicit instances of such maps for the models we discuss.
While the map defines an isomorphism between the \textit{global} operator algebras of the continuum and lattice theories, it preserves locality up to a bounded spread: \textit{local} operators supported in a given region are mapped to operators supported in a slightly enlarged region.

We demonstrate this QFT/Lattice map first for a non-chiral Dirac fermion and then for $N$ non-chiral Majorana fermions in 1+1 dimensions. The resulting lattice Hamiltonian realizes the full non-abelian $\mathrm{O}(N)_\mathrm{L}\times \mathrm{O}(N)_\mathrm{R}$ chiral global symmetry and its 't Hooft anomalies exactly, including the vector and axial $\mathrm{U}(1)$ symmetries of the Dirac fermion as a special case when $N=2$. 
We also provide a new lattice formulation of the Schwinger model using this map and numerically reproduce the mass gap.

Our map is reminiscent of perfect lattice actions \cite{Hasenfratz:1993sp,Bietenholz:1995cy,Bietenholz:1995kr,Bietenholz:1996qc},
which also reproduce the continuum spectrum without cutoff effects.  
The resulting Euclidean spacetime action is only quasi-local, with couplings decaying exponentially in spacetime.
In contrast, our construction is Hamiltonian with nearest-neighbor couplings.

\section{Dirac fermions}

\subsection{The Hamiltonian}

Consider a 1d lattice of $L$ sites.
On every site $j$, there are infinitely many complex fermions $c_j^{(n)}$, labeled by a flavor index $n\in \mathbb{Z}$. They obey
\begin{equation}
\begin{aligned}
&\{c_j^{(n)},c_{j'}^{(m)\dagger}\} = \delta_{j,j'}\delta_{n,m},~~\\
&\{c_j^{(n)},c_{j'}^{(m)}\} =
\{c_j^{(n)\dagger},c_{j'}^{(m)\dagger}\} =0\,.
\end{aligned}\label{eq:CAR}
\end{equation}
Consider the nearest-neighbor  Hamiltonian
\begin{equation}
\begin{aligned}
&H_\text{Dirac}=\sum_{\substack{j=1\\
n\in \mathbb{Z}}}^L
\pi\Bigl(n+\frac12\Bigr)
:c^{(n)\dagger}_jc^{(n)}_j:
\\
&-\frac{\pi}{4}\sum_{\substack{j=1\\n,m\in \mathbb{Z}}}^L\Bigl[(-1)^m\,
W_{m-n}
\,c^{(n)\dagger}_{j+1}c^{(m)}_j
+\mathrm{h.c.}\Bigr],
\label{eq:diracH}
\end{aligned}
\end{equation}
where $W_n = \left(
{\sin( \pi n/4) \over \pi n/4}\right)^2$ and $W_0=1$.  Here $:c_j^{(n)\dagger}c_j^{(n)}:=c_j^{(n)\dagger}c_j^{(n)}$ for $n\ge0$ and $-c_j^{(n)}c_j^{(n)\dagger}$ for $n\le -1$.
We impose $c_{j+L}^{(n)} = e^{2\pi i \nu} c_j^{(n)}$ with $\nu=0$ ($\nu=1/2$) for the periodic (antiperiodic)  boundary condition.

In momentum space, $c_j^{(n)}=L^{-\frac12}\sum_k e^{ikj}c_{k}^{(n)}$, where $
k=\frac{2\pi(p+\nu)}{L}\pmod{2\pi}$ with $p\in\mathbb{Z}$.
The Hamiltonian takes the form
\(H_\text{Dirac}=\sum_k\sum_{n,m}c^{(n)\dagger}_k\,h(k)_{nm}\,c^{(m)}_k\) up to a
constant, with
\begin{equation}
\begin{aligned}
&h(k)_{nm}=\pi\Bigl(n+\frac12\Bigr)\delta_{nm}\\
&-\frac{\pi}{2}(-1)^n\,
W_{m-n}
\times
\begin{cases}\cos k,&m-n\ \text{even},\\ i\sin k,&m-n\ \text{odd}.\end{cases}
\label{eq:dirach}
\end{aligned}
\end{equation}
We claim that diagonalizing this infinite-dimensional matrix yields an \textit{exactly linear} dispersion relation, folded into the periodic Brillouin zone $k\in [-\pi,\pi)$, for any finite system size $L$:
\begin{equation}\label{eq:linear}
E(k) = \pm  (k +2\pi \mathcal{N}) ,~~~\mathcal{N}\in \mathbb{Z}.
\end{equation}
Therefore, this lattice model shares exactly the same spectrum as the field theory of a free, non-chiral, massless Dirac fermion in 1+1d to all energy levels (up to an overall, finite, additive constant proportional to $L$).
Below we derive this lattice model and prove this claim.

\subsection{An exact QFT/Lattice map}

We start with the Hamiltonian for the 1+1d non-chiral massless Dirac fermion field theory,
\begin{equation}
H_\text{Dirac} = i \int dx \left(
\psi_\text{L}^\dagger \partial_x \psi_\text{L}
-\psi_\text{R}^\dagger \partial_x \psi_\text{R}
\right),
\end{equation}
where $\psi_\text{L}(x)$ and $\psi_\text{R}(x)$ are the one-component, left and right Weyl fermion fields, respectively, satisfying $\{\psi_\text{L}(x),\psi_\text{L}^\dagger(x')\}=\{\psi_\text{R}(x),\psi_\text{R}^\dagger(x')\}=\delta(x-x')$, with all other anticommutators vanishing.
This field theory has an $\text{O}(2)_\text{L}\times \text{O}(2)_\text{R}$ global symmetry, where each O(2) acts on a Weyl fermion by a phase and by charge conjugation.

We break the continuous spatial circle into $L$ intervals $[j,j+1]$ and use $0\le u\le 1$ to parametrize each interval. (We have set the lattice spacing to 1 for simplicity.)
We want to find a basis of fields so that each of the new fields is assigned a single interval.
Furthermore, the fields from different intervals are independent of each other.
The naive assignment $\eta_{\text{R},j}(u)\overset{?}{=}  \psi_\text{R}(j+u),\,\eta_{\text{L},j}(u)\overset{?}{=}  \psi_\text{L}(j+u)$ does not satisfy these conditions, because the relation $\eta_{\text{R},j-1}(u=1)=\eta_{\text{R},j}(u=0)$ involves fields from different intervals.

\begin{figure}[h!]
\centering
\begin{tikzpicture}[>=stealth,line width=1.1pt,x=2.6cm,y=1cm,font=\small]
\tikzset{site/.style={dotted,thin,gray}}
\begin{scope}
\node at (-0.45,1.3) {(a)};
\foreach \x/\l in {0/{j-1},1/{j},2/{j+1}}{
  \draw[site] (\x,-0.3)--(\x,1.3); \node[below] at (\x,-0.3) {$\l$};}
\draw[blue] (-0.3,1)--(2.3,1) node[right] {$\psi_\text{R}$};
\draw[blue,->] (0.45,1)--(0.55,1); \draw[blue,->] (1.45,1)--(1.55,1);
\draw[red] (-0.3,0)--(2.3,0) node[right] {$\psi_\text{L}$};
\draw[red,->] (0.55,0)--(0.45,0); \draw[red,->] (1.55,0)--(1.45,0);
\end{scope}
\begin{scope}[yshift=-2.3cm]
\node at (-0.45,1.3) {(b)};
 \foreach \x/\l in 
  {0/{},1/{},2/{}}
 { \draw[site] (\x,-0.3)--(\x,1.3); \node[below] at (\x,-0.3) {$\l$};}
\draw[blue!30] (-0.3,1)--(1,1); \draw[blue!30] (2,1)--(2.3,1);
\draw[blue] (1,1)--(2,1); \draw[blue,->] (1.45,1)--(1.55,1);
\node[blue,above] at (1.5,1.02) {$\psi_\text{R}(j+u)$};
\draw[red!30] (-0.3,0)--(0,0); \draw[red!30] (1,0)--(2.3,0);
\draw[red] (0,0)--(1,0); \draw[red,->] (0.55,0)--(0.45,0);
\draw[red] (1,0.5)--(2,0.5); \draw[red,->] (1.45,0.5)--(1.55,0.5);
\node[red,below] at (1.5,0.48) {$\psi_\text{L}(j-u)$};
\draw[red!70,->,line width=0.7pt] (0.5,0.1) to[bend left=30] (0.97,0.42);
\node[red!70,font=\footnotesize,left] at (0.6,0.42) {fold about $j$};
\fill[red] (1,0.5) circle (1.2pt); \fill[red] (2,0.5) circle (1.2pt);
\node[red,font=\footnotesize,above left] at (1.0,0.55) {$\psi_\text{L}(j)$};
\node[red,font=\footnotesize,right] at (2.02,0.5) {$\psi_\text{L}(j{-}1)$};
\end{scope}
\begin{scope}[yshift=-4.8cm]
\node at (-0.45,1.3) {(c)};
\foreach \x/\l in 
{0/{},1/{},2/{}}
{
  \draw[site] (\x,-0.3)--(\x,1.3); \node[below] at (\x,-0.3) {$\l$};}
\foreach \i in {0,...,23}{
  \pgfmathsetmacro{\ua}{\i/24} \pgfmathsetmacro{\ub}{(\i+1)/24}
  \pgfmathsetmacro{\um}{(\ua+\ub)/2}
  \pgfmathsetmacro{\th}{(\um<=0.5) ? 90-180*\um*\um : 180*(1-\um)*(1-\um)}
  \pgfmathsetmacro{\p}{100*sin(\th)*sin(\th)}
  \pgfmathsetmacro{\q}{100-\p}
  \draw[red!\p!blue] (1+\ua,1)--(1+\ub,1);
  \draw[red!\q!blue] (1+\ua,0.55)--(1+\ub,0.55);
}
\draw[->,red!50!blue] (1.47,1)--(1.53,1); \draw[->,red!50!blue] (1.47,0.55)--(1.53,0.55);
\node[above] at (1.5,1.02) {$\eta_{\text{R},j}(u)$};
\node[below] at (1.5,0.53) {$\eta_{\text{L},j-1}(1-u)$};
\fill[red] (1,1) circle (1.2pt); \fill[blue] (2,1) circle (1.2pt);
\fill[blue] (1,0.55) circle (1.2pt); \fill[red] (2,0.55) circle (1.2pt);
\node[red,font=\footnotesize,left] at (0.98,1) {$\psi_\text{L}(j)$};
\node[blue,font=\footnotesize,right] at (2.02,1) {$\psi_\text{R}(j{+}1)$};
\node[blue,font=\footnotesize,left] at (0.98,0.55) {$-\psi_\text{R}(j)$};
\node[red,font=\footnotesize,right] at (2.02,0.55) {$\psi_\text{L}(j{-}1)$};
\node[font=\footnotesize,gray] at (1,1.35) {$\theta=\pi/2$};
\node[font=\footnotesize,gray] at (2,1.35) {$\theta=0$};
\end{scope}
\begin{scope}[yshift=-7.5cm]
\node at (-0.45,1.3) {(d)};
\foreach \x/\l in {0/{j-1},1/{j},2/{j+1}}{
  \draw[site] (\x,-0.3)--(\x,1.3); \node[below] at (\x,-0.3) {$\l$};}
\foreach \j in {0,1}{
  \foreach \i in {0,...,23}{
    \pgfmathsetmacro{\ua}{0.1+0.8*\i/24} \pgfmathsetmacro{\ub}{0.1+0.8*(\i+1)/24}
    \pgfmathsetmacro{\um}{(\ua+\ub)/2}
    \pgfmathsetmacro{\th}{(\um<=0.5) ? 90-180*\um*\um : 180*(1-\um)*(1-\um)}
    \pgfmathsetmacro{\p}{100*sin(\th)*sin(\th)}
    \draw[red!\p!blue] (\j+\ua,1)--(\j+\ub,1);
    \draw[red!\p!blue] (\j+\ua,0)--(\j+\ub,0);
  }
  \draw[red]  (\j+0.1,0) arc (-90:-270:0.08 and 0.5);
  \draw[blue] (\j+0.9,1) arc (90:-90:0.08 and 0.5);
  \draw[->,red!50!blue] (\j+0.47,1)--(\j+0.53,1);
  \draw[->,red!50!blue] (\j+0.53,0)--(\j+0.47,0);
}
\node[above] at (0.5,1.02) {$\eta_{\text{R},j-1}$}; \node[below] at (0.5,-0.02) {$\eta_{\text{L},j-1}$};
\node[above] at (1.5,1.02) {$\eta_{\text{R},j}$};   \node[below] at (1.5,-0.02) {$\eta_{\text{L},j}$};
\fill[red] (1.02,0.5) circle (1.2pt); \fill[blue] (1.98,0.5) circle (1.2pt);
\node[font=\footnotesize,red,above] at (1.02,1.3) {$\psi_\text{L}(j)$};
\node[font=\footnotesize,blue,above] at (1.98,1.3) {$\psi_\text{R}(j{+}1)$};
\end{scope}
\end{tikzpicture}
\caption{The field redefinition in  \eqref{eq:fold} around site \(j\).
(a) The original right-moving  field \(\psi_\text{R}\) and 
left-moving field \(\psi_\text{L}\). 
(b) Fold: the piece of  \(\psi_\text{L}\) on the interval 
\([j-1,j]\) is folded back about the site \(j\) onto the piece of 
\(\psi_\text{R}\) on \([j,j+1]\), the point \(j-u\) mapped to 
\(j+u\).
(c) Rotate: the new fields $\eta_{\text{L,R}}$ are defined by mixing the $\psi_{\text{L,R}}$  by the angle \(\theta(u)\), which is \(\pi/2\) at the site \(j\) (the
two fields are exchanged) and \(0\) at \(j+1\) (nothing happens).
 (d) Unfold the lower wire back to \([j-1,j]\)
and reconnect the wires following \eqref{eq:matching}.}
\label{fig:wires}
\end{figure}
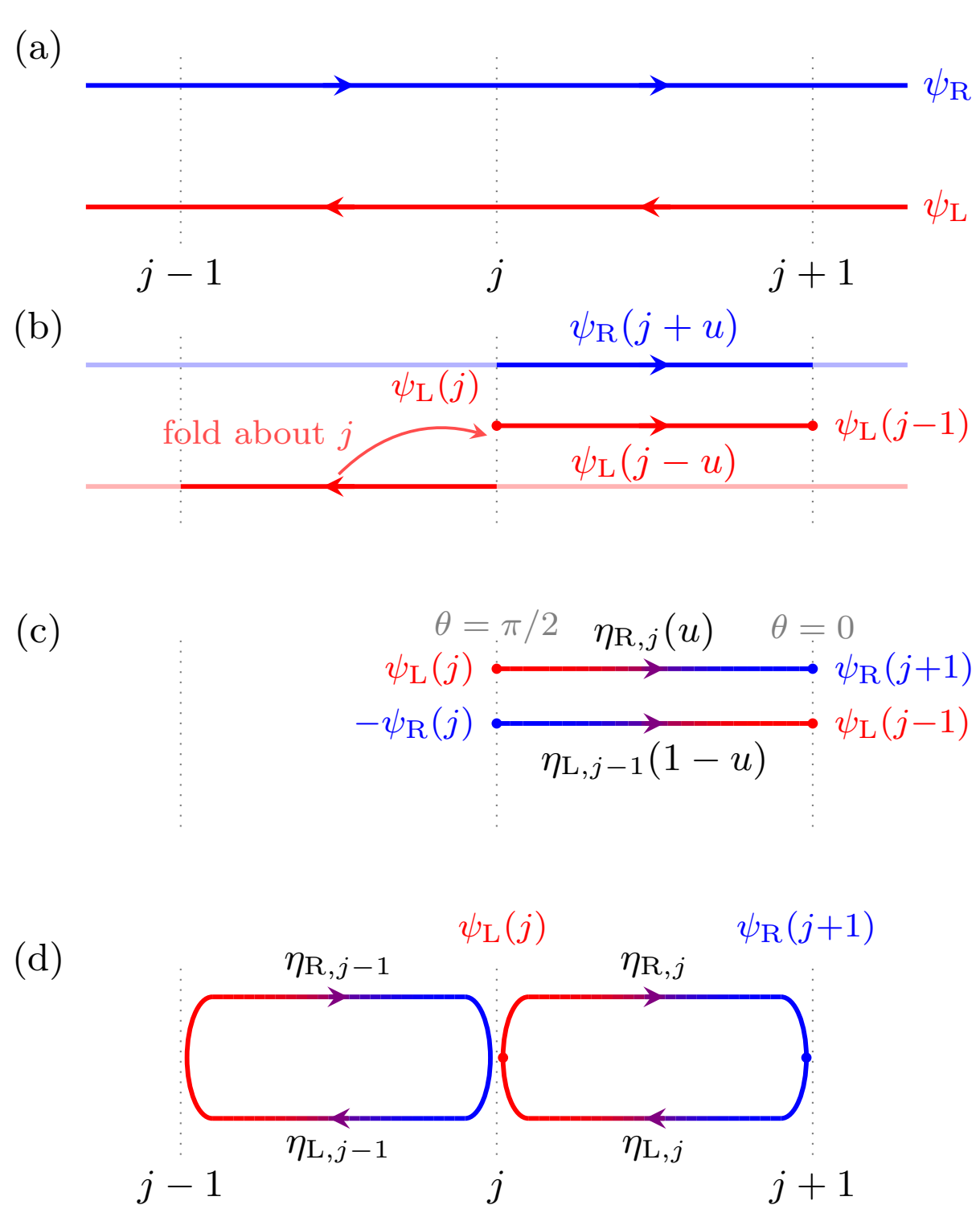

To remedy this, we introduce new fields that mix the right mover to the right of each site $j$ with the left mover to its left, by a rotation whose angle decreases from $\pi/2$ at the site to 0 at the next site.
Explicitly, the new fields, denoted by $\eta_{\text{R},j}(u),\eta_{\text{L},j}(u)$, are defined as follows
\begin{equation}
\begin{pmatrix}\eta_{\text{R},j}(u)\\ \eta_{\text{L},j-1}(1-u)\end{pmatrix}
=\begin{pmatrix}\cos\theta(u)&\sin\theta(u)\\-\sin\theta(u)&\cos\theta(u)\end{pmatrix}
\begin{pmatrix}\psi_\text{R}(j+u)\\ \psi_\text{L}(j-u)\end{pmatrix},
\label{eq:fold}
\end{equation}
where $\theta(u)$ is any function that interpolates from $\theta(0)=\pi/2$ to $\theta(1)=0$. See Fig.\ \ref{fig:wires}.
The continuum Hamiltonian written in terms of the new fields takes the form $H_\text{Dirac}=H_{\mathrm{site}}+H_{\mathrm{link}}$ where
\begin{equation}
\begin{aligned}
H_{\mathrm{site}}&=i\sum_{j=1}^L\int_0^1 du
\bigl(\eta^\dagger_{\text{L},j}\partial_u\eta_{\text{L},j}-\eta^\dagger_{\text{R},j}\partial_u\eta_{\text{R},j}\bigr),\\
H_{\mathrm{link}}&=
i\sum_{j=1}^L\int_0^1 du\,
\theta'(u)\,\Bigl[\eta^\dagger_{\text{R},j}(u)\,\eta_{\text{L},j-1}(1-u)\\
&\qquad\qquad\qquad\quad-\eta^\dagger_{\text{L},j-1}(1-u)\,\eta_{\text{R},j}(u)\Bigr]\,.
\label{eq:Heta}
\end{aligned}
\end{equation}

The new fields $\eta$ from different intervals $[j,j+1]$ are independent of each other and they obey the boundary conditions:
\begin{equation}
\eta_{\text{R},j}(0)=\eta_{\text{L},j}(0),\quad
\eta_{\text{R},j}(1)=-\eta_{\text{L},j}(1) \,.
\label{eq:matching}
\end{equation}
Next, we decompose the new fields into a discrete, complete set of modes (which become the different flavors of lattice fermions, labeled by the index $n$).
The appropriate Fourier expansion subject to the boundary conditions in  \eqref{eq:matching} is\footnote{This can be seen by  unfolding the interval into a circle of length 2 parametrized by $-1\le u\le 1$. We group the two fields into a single one $\xi_j(u) = \eta_{\text{R},j}(u)$ for $0\le u \le 1$ and $\xi_j(u) = \eta_{\text{L},j}(-u)$ for $-1\le u\le0$.
From \eqref{eq:matching}, $\xi_j(u)$ is continuous at $u=0$ and obeys the antiperiodic boundary condition at $u=1$, i.e., $\xi_j (1) = -  \xi_j (-1)$. }
\begin{equation}
\begin{aligned}
\eta_{\text{R},j}(u)&=\frac1{\sqrt2}\sum_{n\in\mathbb Z}
e^{\,i\pi(n+\frac12)(u-\frac12)}\,c^{(n)}_j,\\
\eta_{\text{L},j}(u)&=\frac{-i}{\sqrt2}\sum_{n\in\mathbb Z}
(-1)^n e^{-i\pi(n+\frac12)(u-\frac12)}\,c^{(n)}_j .
\end{aligned}
\label{eq:diraceta}
\end{equation}
The continuum Hamiltonian in \eqref{eq:Heta} written in terms of these Fourier modes $c_j^{(n)}$ then gives the lattice Hamiltonian in \eqref{eq:diracH} with
\begin{equation}
W_n = -{2\over\pi}\int_0^1 du\,\theta'(u)\,e^{i \pi n(u-1/2)}.
\end{equation}
The details of the function $\theta(u)$ do not matter as long as it interpolates from $\pi/2$ to 0.
To simplify some expressions below, we  assume  $\theta(1-u)+\theta(u)=\pi/2$ so that $W_n$ is real.
For concreteness, we can choose
\begin{equation}
\theta(u)
=\begin{cases}
\dfrac{\pi}{2}-\pi u^2,&0\leq u\leq\tfrac12,\\[6pt]
\pi(1-u)^2,&\tfrac12\leq u\leq1,
\end{cases}
\label{eq:theta}
\end{equation}
which gives the $W_n$ below \eqref{eq:diracH}.
We emphasize that the lattice Hamiltonian is an exact rewriting of the continuum one, just in a different basis.
Therefore, the spectra of the lattice and the continuum Hamiltonians match to all energy levels.

\begin{figure}
\centering
\includegraphics[width=\columnwidth]{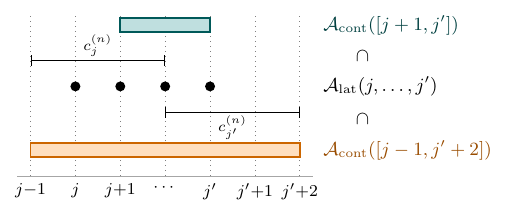}
\caption{The type I lattice algebra is sandwiched between the type III continuum algebras of nested regions.}\label{fig:split}
\end{figure}

We have established an exact  map between the continuum fields $\psi_\text{L,R}$ and the lattice fields $c_j^{(n)}$:
\begin{equation}
\begin{aligned}
c^{(n)}_j=\frac1{\sqrt2}
\int_0^1 d u\,\Bigl[
\cos\theta(u)e^{-i\pi(n+\frac12)(u-\frac12)}\, &\psi_{\text{R}}(j+u)\\
-i\cos\theta(u)(-1)^n e^{\,i\pi(n+\frac12)(u-\frac12)}\,&\psi_{\text{R}}(j+2-u)\\
+\sin\theta(u)
e^{-i\pi(n+\frac12)(u-\frac12)}\, &\psi_{\text{L}}(j-u)\\
+i\sin\theta(u)(-1)^n e^{\,i\pi(n+\frac12)(u-\frac12)}\, &\psi_{\text{L}}(j+u)\, \Bigr],
\end{aligned}
\label{eq:cfrompsi}
\end{equation}
where a lattice field at site $j$ is mapped to continuum fields in the interval $[j-1,j+2]$.
Conversely, for $0\le u<1$,
\begin{equation}
\begin{aligned}
&\psi_{\text{R}}(j+u)=\frac1{\sqrt2}\times\\
&\sum_{n\in\mathbb Z}
e^{\,i\pi(n+\frac12)(u-\frac12)}\Bigl(\cos\theta(u)\,c^{(n)}_j+i(-1)^n\sin\theta(u)\,c^{(n)}_{j-1}\Bigr),\\
&\psi_{\text{L}}(j-u)=\frac1{\sqrt2}\times\\
&\sum_{n\in\mathbb Z}
e^{\,i\pi(n+\frac12)(u-\frac12)}\Bigl(\sin\theta(u)\,c^{(n)}_j-i(-1)^n\cos\theta(u)\,c^{(n)}_{j-1}\Bigr).
\end{aligned}
\label{eq:psifromc}
\end{equation}
These two maps extend to a bounded-spread isomorphism between the global continuum and lattice  operator algebras. 
They map the type III continuum  algebra 
${\cal A}_\text{cont}(I)$ associated with an interval $I$ to the type I$_\infty$ lattice algebra associated with an enlarged interval ${\cal A}_\text{lat}(j,\cdots ,j')$, and vice versa. 
More specifically, we have (see Fig.\ \ref{fig:split}):
\begin{equation}
{\cal A}_\text{cont}([j+1, j'])\subset
{\cal A}_\text{lat}(j,\cdots,j')
\subset
{\cal A}_\text{cont}([j-1, j'+2]),
\end{equation}
which is precisely the split property \cite{Doplicher:1984zz,Buchholz:1986bg,Buchholz:1986dy,Fewster:2016mzz} of the continuum algebra: the lattice algebra provides an explicit type I factor sandwiched between the continuum operator algebras for the intervals $[j+1,j']$ and $[j-1,j'+2]$. 
Since the map is the same for every $j$, lattice translation by one site coincides with continuum translation by one lattice spacing.

It is natural to ask if the flavor index $n\in \mathbb{Z}$ of our lattice fermion $c_j^{(n)}$ can be interpreted as an extra spatial dimension, similar to the picture in  \cite{Shamir:1993zy}.
Put differently, is our model the same as the domain wall fermion construction \cite{Kaplan:1992bt}?
The answer is \textit{no}, and we provide two explanations in the following  subsections.

\subsection{No new massless modes from truncation}\label{sec:truncation}

\begin{figure}
\centering
\includegraphics[width=\columnwidth]{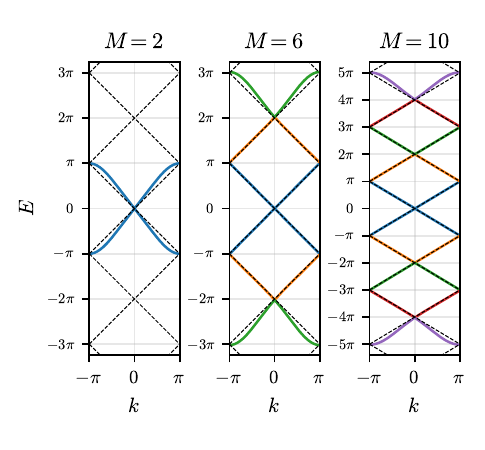}
\caption{The one-particle dispersion relation $E(k)$ for the truncated lattice Hamiltonian in \eqref{eq:diracH} by keeping $M$ fermions per site. As $M$ becomes larger and larger, the dispersion relation becomes closer to the continuum linear dispersion relation  in \eqref{eq:linear} (dashed lines). Unlike the domain wall fermion construction, finite truncation in our model does not introduce extra zeros. }\label{fig:dispersion}
\end{figure}

In the domain wall fermion construction, any truncation of the extra dimension will necessarily introduce extra massless modes.
In contrast, there is no extra massless mode in the truncated Hamiltonian \eqref{eq:diracH} if we
restrict $-{M\over2}\le n,m \le {M\over2}-1$ to keep only $M$ fermions per site.\footnote{When $M$ is a multiple of 4, the truncation lifts the massless modes at $k=0$ by a small gap.  When $M\equiv 2$ mod 4, the massless modes at $k=0$ remain exactly gapless. We will focus on the latter case henceforth.}
For instance, the truncated Hamiltonian with $M=2$ complex fermions per site has the dispersion relation
$E_\pm^{(M=2)} =\pm \sqrt{{\pi^2\over 4}(1-\cos k)^2+{16\over \pi^2}\sin^2k}$, which has a single zero.
In Fig.\ \ref{fig:dispersion}, we plot the dispersion relation for $M=2,6,10$, where it is clear that there is only one left-moving and one right-moving massless mode at $k=0$.

\subsection{Finiteness of entanglement entropy and Hilbert space factorization}\label{sec:entanglement}

Another reason why our model can be treated as a genuine 1+1d model in contrast to the domain wall realization is that the entanglement entropy of any interval in the ground state is finite. 
In particular, it implies that the operator algebras generated by the fermion modes at a given site are all of type I. Hence, the global Hilbert space of the vacuum sector can be decomposed into a ($\mathbb{Z}_2$-graded) tensor product of Hilbert spaces $\mathcal{H}_j$ associated with the sites of the lattice
\begin{equation}
\mathcal{H}_{\text{global}} =  \widehat{\bigotimes}_{j = 1}^{L} \mathcal{H}_j.
\end{equation}

Below we compute the entropy $S^{(M)}_\ell$ of the $M$ fermions $c_j^{(n)}$ with $-\frac M2 \le n \le\frac M2-1$ inside the interval $[1,\ell]$ on an antiperiodic chain of $L$ sites. Following the general formula for free fermion systems \cite{peschel2003calculation}, the entropy is
\[
\begin{aligned}
S_\ell^{(M)}
=-\operatorname{Tr}\bigl[
&C_\ell^{(M)}\log C_\ell^{(M)}
\\[-2pt]
&+(1-C_\ell^{(M)})
\log(1-C_\ell^{(M)})
\bigr],
\end{aligned}
\]
where
\begin{equation}
\begin{aligned}
C_\ell^{(M)}&
=
\Bigl(
\langle c_j^{(n)\dagger}c_{j'}^{(m)}\rangle
\Bigr)_{\substack{1\leq j,j'\leq\ell\\-\frac M2\leq n,m\leq \frac M2-1}}\\
&=\frac{1}{L}\sum_{k>0}
b\bigl(k+\pi(n+\tfrac12)\bigr)\,b\bigl(k+\pi(m+\tfrac12)\bigr)
\\[-2pt]
&\quad\times
\cos\bigl(\tfrac{\pi}{2}(n-m)+k(j-j')\bigr),
\end{aligned}
\end{equation}
is the restriction of the exact two-point function and the sum in $k$ is over all positive $k\in {2\pi\over L} (\mathbb{Z}+\frac12)$.
Here 
\[
\begin{aligned}
b(z)=2\int_0^{1/2}\!du\,\bigl[
&\cos(\pi u^2)\cos(zu)
\\[-2pt]
&+\sin(\pi u^2)\cos\bigl(z(1-u)\bigr)
\bigr].
\end{aligned}
\]

\begin{figure}[t]
  \centering
  \includegraphics[width=\columnwidth]{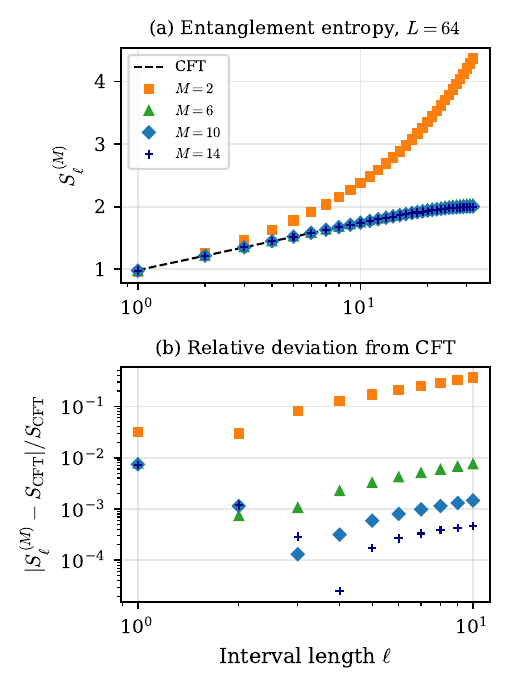}
  \caption{
    Entanglement entropy associated with an interval of length $\ell$ on a circle  for $L=64$ sites with antiperiodic boundary conditions. (a) Entanglement entropy of the first $M=2,6,10,14$ fermions per site and the CFT prediction \eqref{eq:CFTintentanglement}. (b) Relative deviations
    $|S_\ell^{(M)}-S_{\mathrm{CFT}}|/S_{\mathrm{CFT}}$
    on logarithmic axes for $\ell\leq 10$. 
  }
  \label{fig:dirac-interval-entropy}
\end{figure}

We check numerically for $L = 64$ that the limit $S_{\ell} = \lim_{M \to \infty} S^{(M)}_{\ell}$ exists and is in excellent agreement with the formula expected from conformal field theory with $c=1$ \cite{holzhey1994geometric,Calabrese:2004eu}:
\begin{equation}    \label{eq:CFTintentanglement}
S_{\ell} \approx \frac13 \log \left( \frac{L}{\pi} \sin \left( \frac{\pi \ell}{L}\right) \right) + C
\end{equation}
for a constant $C \approx 0.987$ (see Fig.\ \ref{fig:dirac-interval-entropy}).
This finite entanglement entropy stands in contrast to the domain wall fermion, whose entanglement entropy grows linearly in the truncation $M$ because of the area law in 2+1d.

One can also extend our analysis to finite temperature Gibbs states and see that it agrees with the CFT result.

\subsection{Exact vector and axial lattice  symmetries}

Since the lattice Hamiltonian is an exact rewriting of the continuum one, \eqref{eq:diracH} has exactly the same $\text{O}(2)_\text{L}\times \text{O}(2)_\text{R}$ global symmetry as the continuum field theory. 
Let us compare the chiral symmetries realized in other lattice models for a Dirac fermion. 
The fermionic Villain model in \cite{Lu:2026itw,2026arXiv260802722D} only realizes the $[\text{U(1)}_\text{L}\times \text{U(1)}_\text{R}]\rtimes \mathbb{Z}_2^\text{C}$ subgroup faithfully. 
The remaining continuum $\mathbb{Z}_2$ symmetry, which acts on $\psi_\text{L}$ by charge conjugation, is mixed with translation in the fermionic Villain model. 
The staggered fermion model realizes both the vector charge $Q_\text{V}=Q_\text{L}+Q_\text{R}$ and the axial charge $Q_\text{A}=-Q_\text{L}+Q_\text{R}$, but they do not commute and generate the non-abelian Onsager algebra \cite{Chatterjee:2024gje} (see also \cite{Vernier:2018han,Pace:2024oys}). 
The staggered fermion model also realizes a U(1)$_\text{V}\times \mathbb{R}$ global symmetry, where the axial symmetry is generated by a non-compact charge \cite{Creutz:2001wp,Chatterjee:2024gje,Lew-Smith:2026jqb}.

We will focus on the $[\text{U(1)}_\text{V}\times \text{U(1)}_\text{A}]/\mathbb{Z}_2$ subgroup, while the chiral charge conjugation can be realized in a way similar to the chiral $\mathbb{Z}_2$ symmetry of the Majorana fermion in Sec.\ \ref{sec:Z2}. 
In terms of the continuum fields, the vector and axial charges are $Q_\text{V} = \int dx :\psi_\text{R}^\dagger \psi_\text{R}
+\psi_\text{L}^\dagger \psi_\text{L}:$ and $Q_\text{A} = \int dx :\psi_\text{R}^\dagger \psi_\text{R}
-\psi_\text{L}^\dagger \psi_\text{L}:$. 
In terms of the lattice fields, 
the $\text{U(1)}_\text{V}$ symmetry acts on-site as $c_j^{(n)}\to e^{-i\alpha}c_j^{(n)}$. 
The vector charge is simply $Q_\text{V} = \sum_{j=1}^L \sum_{n\in \mathbb{Z}}:c_j^{(n)\dagger} c_j^{(n)}:$. 
The axial charge is not on-site but involves nearest-neighbor couplings:
\begin{equation}
\begin{aligned}
Q_{\text{A}}=i\sum_j\Bigl[&\sum_{\substack{n,m\\ m-n\ \text{odd}}}
C_{m-n}c^{(n)\dagger}_jc^{(m)}_j\\
&+\sum_{n,m}\frac{(-1)^m}{2}S_{m-n}\;
\bigl(c^{(n)\dagger}_jc^{(m)}_{j-1}-c^{(m)\dagger}_{j-1}c^{(n)}_j\bigr)\Bigr],
\end{aligned}
\label{eq:QAmodes}
\end{equation}
where 
\begin{equation}
\begin{aligned}
&C_n=\int_0^1 du\,\cos2\theta(u)\,\sin\bigl(\pi n(u-\tfrac12)\bigr),\\
&S_n=\int_0^1 du\, \sin2\theta(u)\,\cos\bigl(\pi n(u-\tfrac12)\bigr)\,,
\end{aligned}\label{eq:CnSn}
\end{equation}
satisfying $|C_n| \leq 5/|n|^3$ for odd $n$ and $|S_n| \leq 5/|n|^3$. 
Both $e^{i \alpha Q_\text{V}}$ and $e^{i \alpha Q_\text{A}}$ define strongly continuous families of unitary operators acting on the global Hilbert space. By Stone's theorem, both $Q_\text{V}$ and $Q_\text{A}$ give rise to self-adjoint operators.

The U(1)$_\text{A}$ symmetry acts on the fermions as
\begin{equation}
\begin{aligned}
&e^{i\alpha Q_{\text{A}}}\,c^{(n)}_j\,e^{-i\alpha Q_{\text{A}}}
=\cos\alpha\;c^{(n)}_j
+\sin\alpha \; \widetilde c_j^{(n)},
\end{aligned}
\label{eq:axialaction}
\end{equation}
where
\begin{equation}
\begin{aligned}
\widetilde c_j^{(n)}&=\sum_{\substack{m\in\mathbb Z\\m-n\ \mathrm{odd}}}
C_{m-n}\,c_j^{(m)}\\
&+\frac12\sum_{m\in\mathbb Z}S_{m-n}
\left[
(-1)^m c_{j-1}^{(m)}
-(-1)^n c_{j+1}^{(m)}
\right].
\end{aligned}
\end{equation}
Note that $\widetilde c_j^{(n)}$ obeys the canonical anticommutation relations in \eqref{eq:CAR}. 
Clearly, $[Q_\text{V},Q_\text{A}]=0$.\footnote{More precisely, since $Q_\text{V}$ and $Q_\text{A}$ are unbounded operators, we actually mean strong commutativity: self-adjoint operators $Q_\text{V}$ and $Q_\text{A}$ commute if $e^{i \alpha Q_\text{V}} e^{i \beta Q_\text{A}} = e^{i \beta Q_\text{A}} e^{i \alpha Q_\text{V}}$ for any $\alpha, \beta \in \mathbb{R}$. We use the same convention below.}
Our general construction implies that $Q_\text{V}$ and $Q_\text{A}$ have integer spectra and, on the antiperiodic chain, $\exp(i\pi Q_\text{V}) =\exp(i\pi Q_\text{A})$, so they generate an exact [U(1)$_\text{V}\times$ U(1)$_\text{A}]/\mathbb{Z}_2$ chiral global symmetry. 
However, once the flavor index is truncated, $Q_\text{A}$ becomes non-quantized and no longer commutes with the Hamiltonian. The breaking of the axial symmetry is nevertheless mild: it is violated only by the high-energy modes, while the low-energy physics has an approximate axial symmetry.

We can decompose the global charges into sums of local commuting terms
\[
\begin{aligned}
q_j^\text{A}
&=
\frac{i}{2}\sum_{n\in\mathbb Z}
:\!\left(
c_j^{(n)\dagger}\widetilde c_j^{(n)}
-\widetilde c_j^{(n)\dagger}c_j^{(n)}
\right)\!:\,,
\\
q_j^\text{V}
&=
\frac12\sum_{n\in\mathbb Z}
:\!\left(
c_j^{(n)\dagger}c_j^{(n)}
+\widetilde c_j^{(n)\dagger}\widetilde c_j^{(n)}
\right)\!:.
\end{aligned}
\]
Each $q^\text{V}_j$ and $q^\text{A}_j$ defines a self-adjoint operator that exponentiates to a unitary operator $e^{i \alpha q^{\text{V},\text{A}}_j}$ supported at three sites. The 't Hooft anomaly between U(1)$_\text{V}$ and U(1)$_\text{A}$ can be seen from their equal-time commutators
\[
\begin{aligned}
[q_j^\text{A},q_l^\text{A}]=[q_j^\text{V},q_l^\text{V}]=0,\\
[q_j^\text{V},q_l^\text{A}]
=
\frac{i}{2\pi}
\left(\delta_{l,j-1}-\delta_{l,j+1}\right)\mathbf{1},
\end{aligned}
\] 
corresponding to the Schwinger term \cite{Schwinger:1959xd}. Note that both $e^{i \alpha Q_\text{V}}$ and $e^{i \alpha Q_\text{A}}$ can be represented as quantum circuits consisting of commuting gates.

\section{Schwinger model}

In this section we build an exact lattice model for  U(1) gauge theory coupled to a Dirac fermion, i.e., the  Schwinger model \cite{Schwinger:1962tp}. We start by assuming that space is an infinite line. The continuum Hamiltonian is
\[
\begin{aligned}
H_{\text{Schwinger}} ={}&  i\!\int\! dx \bigl(
\psi_\text{L}^\dagger (\partial_x-i e A) \psi_\text{L}
-\psi_\text{R}^\dagger (\partial_x-i e A) \psi_\text{R}
\bigr) \\
&+ \frac{e^2}{2} \int dx\, E(x)^2
\end{aligned}
\]
subject to the Gauss law  constraint $\partial_x E = \rho_\text{V}$, where $[E(x), e A(y)] = i\delta(x-y)$.

The lattice Hamiltonian can be obtained by covariantizing the QFT/Lattice map:
\begin{equation}
\begin{aligned}
\psi_{\text{R}}(j+u)={}&\frac1{\sqrt2}\sum_{n\in\mathbb Z}e^{i\pi(n+\frac12)(u-\frac12)}\\
&\times\Bigl(\cos\theta(u)\,W\bigl(j+u,j\bigr)\,c^{(n)}_j\\
&\quad+i(-1)^n\sin\theta(u)\,W\bigl(j+u,j-1\bigr)\,c^{(n)}_{j-1}\Bigr),\\
\psi_{\text{L}}(j-u)={}&\frac1{\sqrt2}\sum_{n\in\mathbb Z}e^{i\pi(n+\frac12)(u-\frac12)}\\
&\times\Bigl(\sin\theta(u)\,W\bigl(j-u,j\bigr)\,c^{(n)}_j\\
&\quad-i(-1)^n\cos\theta(u)\,W\bigl(j-u,j-1\bigr)\,c^{(n)}_{j-1}\Bigr),
\end{aligned}
\label{eq:psifromcMOD}
\end{equation}
where
\begin{equation}
W(x,y)=\exp\Bigl(-ie\int_x^y A(z)\,dz\Bigr)
\end{equation}
is the Wilson line.

We define the lattice gauge fields in terms of the continuum ones. 
The lattice holonomy field, which is a unitary operator living on the link, is defined in terms of the Wilson line:
\begin{equation}
U_{j,j+1} = W(j,j+1)
\end{equation}
and  $U_{j,j'}=U_{j,j+1}U_{j+1,j+2}\cdots U_{j'-1,j'}$ for $j<j'$, with $U_{j',j}=U_{j,j'}^\dagger$ and $U_{j,j} = 1$. The lattice electric field is defined through the integrated Gauss law in infinite space:
\begin{equation}
 E_{j,j+1} = E(x) - \int_{-\infty}^{x} \rho_\text{V}(y)\,dy
+\sum_{\substack{s=-\infty\\
n\in \mathbb{Z}}}^{j} : c^{(n)\dagger}_s c^{(n)}_s :  ,
\label{eq:IntGaussLaw}
\end{equation}
for any $j < x < j+1$; on gauge-invariant states the right-hand side does not depend on the choice of $x$. They satisfy the standard commutation relation \cite{Banks:1975gq}:
\begin{equation}
[E_{j,j+1}, U_{j',j'+1}] = \delta_{j,j'}U_{j,j+1}.
\end{equation}

Now that we have formulated the map from the continuum to the lattice on an infinite line, it is straightforward to write the Hamiltonian in terms of these lattice fields on a circle:
\begin{equation}
\begin{aligned}
&H_\text{Schwinger}=\sum_{\substack{j=1\\
n\in \mathbb{Z}}}^L
\pi\Bigl(n+\frac12\Bigr)
:c^{(n)\dagger}_jc^{(n)}_j:
\\
&-\frac{\pi}{4}\sum_{\substack{j=1\\n,m\in \mathbb{Z}}}^L\Bigl[(-1)^m\,
W_{m-n}
\,c^{(n)\dagger}_{j+1}U^\dagger_{j,j+1}c^{(m)}_j
+\mathrm{h.c.}\Bigr]\\
&+ \frac{e^2}{2} \sum_{j=1}^L\int_0^1\!du\; \left( \mathcal E_j(u)^2 - \langle \mathcal E_j(u)^2 \rangle_0 \right),
\end{aligned}
\label{eq:schwingerH}
\end{equation}
subject to the Gauss law  constraint
\begin{equation}
E_{j,j+1}-E_{j-1,j}=\sum_{n\in\mathbb Z}{:}c^{(n)\dagger}_jc^{(n)}_j{:}\,.
\end{equation}
Here
\begin{equation}
\begin{aligned}
&\mathcal E_j(u)=E_{j,j+1}\\
&+\sum_{a,b=-1}^{1}\sum_{n,m\in\mathbb Z}X^{ab}_{nm}(u)\,{:}c^{(n)\dagger}_{j+a}\,U_{j+a,j+b}\,c^{(m)}_{j+b}{:}\,
\end{aligned}
\end{equation}
is the continuum electric field $\mathcal E_j(u) = E(j+u)$, $0\le u<1$, obtained via \eqref{eq:IntGaussLaw}, while $\langle \mathcal E_j(u)^2 \rangle_0$ is the formal expectation value in the Fock vacuum\footnote{
The Fock vacuum is the vector in the global Hilbert space annihilated by $c^{(n \geq 0)}_j$, $c^{(n< 0)\dagger}_j$ and with eigenvalue $0$ for all $E_{j,j+1}$.
} which is subtracted for finiteness of energy. More precisely,
\begin{equation}
\begin{aligned}
&\mathcal E_j(u)^2-\langle\mathcal E_j(u)^2\rangle_0={:}\mathcal E_j(u)^2{:}\\
&\;+\sum_{a,b,c=-1}^{1}\sum_{n,m,k\in\mathbb Z}\operatorname{sgn}\bigl(k+\tfrac12\bigr)X^{ac}_{nk}(u)X^{cb}_{km}(u)\\
&\qquad\times{:}c^{(n)\dagger}_{j+a}U_{j+a,j+b}\,c^{(m)}_{j+b}{:}\,.
\end{aligned}
\end{equation}
The explicit expressions for $X^{ab}_{nm}(u)$ are obtained by expressing $\rho_\text{V}(x)$ in terms of fermionic modes $c^{(n)}_j$, $c^{(n)\dagger}_j$. The non-vanishing $X^{ab}_{nm}(u)$ are
\begin{equation}
\begin{aligned}
X^{1,1}_{nm}(u)&=\frac12\int_0^u\!dv\,\cos^2\theta(v)\,e^{-i\pi(m-n)(v-\frac12)},\\
X^{1,0}_{nm}(u)&=-\frac{i(-1)^m}4\int_0^u\!dv\,\sin2\theta(v)\,e^{-i\pi(m-n)(v-\frac12)},\\
X^{0,1}_{nm}(u)&=\frac{i(-1)^n}4\int_0^u\!dv\,\sin2\theta(v)\,e^{-i\pi(m-n)(v-\frac12)},\\
X^{-1,-1}_{nm}(u)&=-\frac{(-1)^{n+m}}2\int_u^1\!dv\,\sin^2\theta(v)\,e^{i\pi(m-n)(v-\frac12)},\\
X^{0,-1}_{nm}(u)&=-\frac{i(-1)^m}4\int_u^1\!dv\,\sin2\theta(v)\,e^{i\pi(m-n)(v-\frac12)},\\
X^{-1,0}_{nm}(u)&=\frac{i(-1)^n}4\int_u^1\!dv\,\sin2\theta(v)\,e^{i\pi(m-n)(v-\frac12)},\\
X^{0,0}_{nm}(u)&=-\frac12\int_u^1\!dv\,\Bigl[\cos^2\theta(v)\,e^{i\pi(m-n)(v-\frac12)}\\
&\qquad+(-1)^{n+m}\sin^2\theta(v)\,e^{-i\pi(m-n)(v-\frac12)}\Bigr]\\
&\quad-\frac{(-1)^{n+m}}2\int_0^1\!dv\,\sin^2\theta(v)\,e^{i\pi(m-n)(v-\frac12)}.
\end{aligned}
\end{equation}
The formal expression for $H_{\text{Schwinger}}$ defines a self-adjoint operator on the Hilbert space 
\begin{equation}
\mathcal{H}_{\text{global}} \otimes \biggl( \bigotimes_{j = 1}^{L} \mathcal{H}^{(\text{rotor})}_{j,j+1}    \biggr)
\end{equation}
commuting with the Gauss law constraints. The vectors obtained by applying polynomials in creation and annihilation operators $c^{(n)}_j$, $c^{(n)\dagger}_j$, $U_{j,j+1}$, $U^{\dagger}_{j,j+1}$ to the Fock vacuum are all in the domain of this operator.

\begin{figure}[t]
\centering
\includegraphics[width=.37\textwidth]{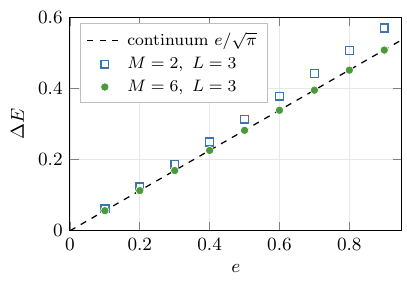}
\caption{The lowest energy gap for system size \(L=3\) and \(M=2,6\) fermions per site, against the exact continuum answer \(e/\sqrt{\pi}\) (dashed). The electric field on the link \(\{L,L+1\}\) is truncated to \(|E_{L,L+1}|\le6\).}
\label{fig:schwinger}
\end{figure}

We numerically simulate the truncated model by keeping only the fermionic modes $c^{(n)}_j$, $c^{(n)\dagger}_j$ with $n=-\frac M2,\dots,\frac M2-1$ on a closed chain with $L$ sites. 
We work in the gauge-invariant subspace, where the Gauss law fixes $E_{j,j+1}=E_{L,L+1}+\sum_{s = 1}^{j}\sum_n{:}c^{(n)\dagger}_sc^{(n)}_s{:}$ for $j<L$, so the only independent rotor is the one on the link $\{L,L+1\}$, and we truncate it to $|E_{L,L+1}|\le E_{\max}$ with $E_{\max}=6$. 
With $M=6$ fermion modes per site on $L=3$ sites and no extrapolation, our gap deviates from the continuum gap $e/\sqrt\pi$ by at most $0.09\%$ for $0.1\le e\le0.9$. See Table~\ref{tab:schwinger} and Fig.~\ref{fig:schwinger}.

For comparison, Hamiltonian lattice formulations with staggered fermions \cite{Banks:1975gq} recover  $e/\sqrt\pi$ only after extrapolating to the continuum and infinite-volume limits. Exact diagonalization of finite staggered lattices followed by extrapolation determined the gap to within $0.7\%$ in \cite{Crewther:1979ka}, and later to within $10^{-6}\%$ in \cite{Cichy:2012rw}. The convergence to the continuum limit is greatly improved by a mass shift arising from the discrete chiral symmetry of the lattice \cite{Dempsey:2022nys}. In our model, by contrast, the lattice Hamiltonian is an exact rewriting of the continuum one, so no such shift is needed. 

\begin{table}[t]
\centering
\begin{tabular}{c|c|c|c}
\(e\) & \(\Delta E\) & \(e/\sqrt{\pi}\) & \(\Delta E\,/\,(e/\sqrt{\pi})\)\\ \hline
\(0.1\) & \(0.0564\) & \(0.0564\) & \(0.9993\)\\
\(0.2\) & \(0.1128\) & \(0.1128\) & \(0.9993\)\\
\(0.3\) & \(0.1691\) & \(0.1693\) & \(0.9993\)\\
\(0.4\) & \(0.2255\) & \(0.2257\) & \(0.9993\)\\
\(0.5\) & \(0.2819\) & \(0.2821\) & \(0.9994\)\\
\(0.6\) & \(0.3383\) & \(0.3385\) & \(0.9994\)\\
\(0.7\) & \(0.3947\) & \(0.3949\) & \(0.9993\)\\
\(0.8\) & \(0.4510\) & \(0.4514\) & \(0.9992\)\\
\(0.9\) & \(0.5073\) & \(0.5078\) & \(0.9991\)
\end{tabular}
\caption{Lowest energy gap \(\Delta E\) of the truncated lattice Hamiltonian \eqref{eq:schwingerH} with $M=6$ fermions per site and system size \(L=3\). The electric field on the link \(\{L,L+1\}\) is truncated to \(|E_{L,L+1}|\le6\); all other link fields are fixed by the Gauss law. In the continuum, the lowest gap of the massless Schwinger model on a circle is exactly \(e/\sqrt{\pi}\) for any system size. The first three columns are in units where the lattice spacing is set to 1.}
\label{tab:schwinger}
\end{table}

\section{Majorana fermions}

\subsection{The Hamiltonian}

Consider the Hamiltonian for the 1+1d non-chiral massless Majorana fermion field theory,
\begin{equation}
H = {i\over2} \int dx \left(
\chi_\text{L}\partial_x \chi_\text{L}
-
\chi_\text{R}\partial_x \chi_\text{R}
\right),
\end{equation}
where $\chi_\text{L}(x)$ and $\chi_\text{R}(x)$ are the left and the right Majorana--Weyl fermion fields, respectively, satisfying $\{\chi_\text{L}(x),\chi_\text{L}(x')\}=\{\chi_\text{R}(x),\chi_\text{R}(x')\}=\delta(x-x')$, $\{\chi_\text{L}(x),\chi_\text{R}(x')\}=0$.
Following the same map in \eqref{eq:fold} with $\psi$ replaced by $\chi$, we can write this continuum Hamiltonian in terms of the lattice fields (see App.\ \ref{app:majorana} for the explicit map).
On every site $j$, there are infinitely many complex fermions $c_j^{(r)}$, labeled by a flavor index $r\in \mathbb{Z}$ and obeying the canonical anticommutation relations \eqref{eq:CAR}.
The lattice Hamiltonian is
\begin{equation}
\begin{aligned}
&H =  \sum_{\substack{j=1 \\ r\in \mathbb{Z}}}^L
2\pi\left(r-\frac14\right)
:c_j^{(r)\dagger}c_j^{(r)}:\\
&
+\frac{\pi}{4}
\sum_{\substack{j=1\\r,s\in \mathbb{Z}}}^L
\left.\left(
W_{2(r-s)}c_{j+1}^{(r)\dagger}c_j^{(s)}
+W_{2(r+s)-1} c_{j+1}^{(r)}c_j^{(s)}
+\text{h.c.}\right.\right)
\end{aligned}\label{eq:Hlat}
\end{equation}
with the same $W_n$ as in \eqref{eq:diracH}.  
Here $:c_j^{(r)\dagger}c_j^{(r)}:=c_j^{(r)\dagger}c_j^{(r)}$ for $r\ge1$ and $-c_j^{(r)}c_j^{(r)\dagger}$ for $r\le 0$.

In momentum space, $c_j^{(r)}=L^{-\frac12}\sum_k e^{ikj}c_{k}^{(r)}$, where, as for the Dirac fermion, $k=\frac{2\pi(p+\nu)}{L}\pmod{2\pi}$ with $p\in\mathbb{Z}$, and $\nu=0$ ($\nu=\frac12$) for the periodic (antiperiodic) boundary condition. 
The Hamiltonian takes the form $H=\frac 12\sum_{k,r,s}\Psi_{k,r}^\dagger h(k)_{rs} \Psi_{k,s}$ where $\Psi_{k,r} = (c_k^{(r)} ~~c_{-k}^{(r)\dagger})^T$,
and
\begin{equation}
\begin{aligned}
h(k)_{rs}&=
\left[
\Bigl(2\pi r-\frac{\pi}{2}\Bigr)\delta_{rs}
+\frac{\pi}{2}\,W_{2(r-s)}\cos k\right]\sigma^z\\
&-\sigma^y
\frac{\pi}{2}\,W_{2(r+s)-1}\sin k.
\end{aligned}
\end{equation}
The general argument implies that the spectrum with all flavors
included is exactly \(\{k+2\pi\mathcal{N}\}\cup\{-(k+2\pi\mathcal{N})\}\),
\(\mathcal{N}\in\mathbb Z\), the same  linear dispersion relation as the continuum field theory.

The discussions in Secs.\ \ref{sec:truncation} and \ref{sec:entanglement} apply to the Majorana lattice model as well.
If we restrict $-{M-1\over2}\le r,s \le {M-1\over2}$ to keep only $M$ fermions per site with $M$ odd, the one-particle dispersion relation coincides with that of the truncated Dirac model with $2M$ fermions per site shown in Fig.\ \ref{fig:dispersion}.
For instance, the truncated Majorana Hamiltonian with $M=1$ complex fermion per site has the same dispersion relation as the truncated Dirac Hamiltonian with $M=2$, i.e., $E_\pm =\pm \sqrt{{\pi^2\over 4}(1-\cos k)^2+{16\over \pi^2}\sin^2k}$, which has a single zero.
Similarly, the entanglement entropy of the first $M$ modes inside an interval is half of the corresponding Dirac one with $2M$ modes, and approaches \eqref{eq:CFTintentanglement} with $\frac13$ replaced by $\frac16$ and with $C\approx 0.494$.

\subsection{Chiral $\mathbb{Z}_2$ as an involutive fermionic QCA}\label{sec:Z2}

 The continuum Majorana fermion field theory has a chiral $\mathbb{Z}_2$ global symmetry transformation $\chi_\text{L}\to -\chi_\text{L}$, $\chi_\text{R}\to \chi_\text{R}$. 
In the critical Kitaev chain \cite{Kitaev:2000nmw}, which is also a  lattice regularization of the free massless Majorana fermion, this symmetry is realized by translation by one Majorana site \cite{Seiberg:2023cdc,Seiberg:2025zqx}. More generally, its lattice realization is a nontrivial fermionic quantum cellular automaton (QCA), which cannot be implemented by an involutive operator $U$ (i.e., $U^2=1$) on a lattice with finite-dimensional local Hilbert spaces \cite{2019arXiv190205957J}. However, an involutive realization of this fermionic QCA has recently been constructed in a lattice model with infinite-dimensional local Hilbert spaces \cite{wilburZ2action}.

Using the exact QFT/Lattice maps in \eqref{eq:cfromchi} and \eqref{eq:chifromc}, we can also realize this chiral $\mathbb{Z}_2$ symmetry in our lattice model as an involutive transformation:
\begin{equation}
\begin{aligned}
c_{j}^{(r)}\mapsto &-i\sum_s c^{(s)\dagger}_{j}
C_{2(r+s)-1}
+\frac i2\sum_s\bigl(c_{j+1}^{(s)}-c_{j-1}^{(s)}\bigr)S_{2(r-s)}\\
&+\frac i2\sum_s\bigl(c^{(s)\dagger}_{j+1}+c^{(s)\dagger}_{j-1}\bigr)S_{2(r+s)-1} ,
\end{aligned}
\label{eq:z2modes}
\end{equation}
where $C_n$ and $S_n$ are given in \eqref{eq:CnSn}. This transformation extends to a unitary operator acting on the global Hilbert space $\mathcal{H}_{\text{global}}$. It realizes the nontrivial fermionic QCA in 1+1d in \cite{Jones:2026gmd} as an involutive map.

\section{Chiral SO$(N)$ symmetries}

Our lattice construction also gives rise to exact non-abelian chiral global symmetries. 
Take $N$ copies of the Majorana lattice model \eqref{eq:Hlat}. 
The lattice fermions are denoted as
\(c^{(r)}_{I,j}\) with \(I=1,\ldots,N\), one copy per continuum field
\(\chi^I_{\text{L},\text{R}}\). 
The continuum theory has the symmetry
\(\text{O}(N)_{\text{L}}\times \text{O}(N)_{\text{R}}\), rotating the left and  right
movers independently. 
The continuum generators are, for \(I\neq J\),
\begin{equation}
Q_{\text{L}}^{IJ}=-i\int dx \chi^I_{\text{L}}\chi^J_{\text{L}}\,,\qquad
Q_{\text{R}}^{IJ}=-i\int dx\chi^I_{\text{R}}\chi^J_{\text{R}}\,.
\label{eq:QLcont}
\end{equation}
In terms of the lattice fields, 
the diagonal \(\text{SO}(N)_{\text{V}}\) is on-site and is generated by
\begin{equation}
Q_{\text{V}}^{IJ} = Q_{\text{R}}^{IJ}+Q_{\text{L}}^{IJ}=-i\sum_{j,r}\Bigl(c^{(r)\dagger}_{I,j}c^{(r)}_{J,j}-c^{(r)\dagger}_{J,j}c^{(r)}_{I,j}\Bigr).
\label{eq:QVson}
\end{equation}
On the other hand, the axial charges are non-on-site:
\begin{equation}
\begin{aligned}
&Q_{\text{A}}^{IJ}=Q_{\text{R}}^{IJ}-Q_{\text{L}}^{IJ}=\sum_j\sum_{r,s}\Bigl\{C_{2(r+s)-1}\Bigl(c^{(r)}_{I,j}c^{(s)}_{J,j}+\mathrm{h.c.}\Bigr)\\
&
+\frac12S_{2(r-s)}\Bigl[\Bigl(c^{(r)\dagger}_{I,j}c^{(s)}_{J,j+1}-c^{(r)\dagger}_{I,j+1}c^{(s)}_{J,j}\Bigr)+\mathrm{h.c.}\Bigr]\\
&-\frac12S_{2(r+s)-1}\Bigl[\Bigl(c^{(r)}_{I,j}c^{(s)}_{J,j+1}+c^{(r)}_{I,j+1}c^{(s)}_{J,j}\Bigr)+\mathrm{h.c.}\Bigr]\Bigr\}.
\end{aligned}
\label{eq:QAson}
\end{equation}
These charges commute with the Hamiltonian and exactly realize the $\mathfrak{so}(N)_\text{L}\times \mathfrak{so}(N)_\text{R}$ Lie algebra, i.e., $[Q_{\text{L}}^{IJ},Q_{\text{L}}^{JK}]=-iQ_{\text{L}}^{IK}$, $[Q_{\text{R}}^{IJ},Q_{\text{R}}^{JK}]=-iQ_{\text{R}}^{IK}$, and $[Q_{\text{L}}^{IJ},Q_{\text{R}}^{KM}]=0$. 

Similarly to the vector and axial U(1) symmetries of the Dirac fermion, one can represent the group elements of $\mathrm{SO}(N)_\text{L} \times \mathrm{SO}(N)_\text{R}$ by quantum circuits acting on our lattice system.

\section{Outlook}

The results of \cite{Jones:2026gmd} suggest that such an exact QFT/Lattice map, using infinite-dimensional local Hilbert space, exists whenever the QFT has no pure gravitational anomaly, including the Standard Model. 
Important future directions include generalizing these maps to higher spacetime dimensions, disentangling anomaly-free chiral symmetries using infinite-dimensional ancillary degrees of freedom as in \cite{Thorngren:2026ydw,Gioia:2026wdl}, and building lattice chiral gauge theories. These will be presented in upcoming works.

\section*{Acknowledgements}

We thank Shailesh Chandrasekharan, Ross Dempsey, Lei Gioia, Jake McNamara, Nathan Seiberg, Sahand Seifnashri, Wilbur Shirley, Tin Sulejmanpasic, Silviu Pufu, and Ryan Thorngren for discussions. 
S.H.S. was supported in part by the Simons Collaboration on Ultra-Quantum Matter, which is a grant from the Simons Foundation (651444,  SHS), in part by the U.S. Department of Energy, Office of Science, Office of High Energy Physics of U.S. Department of Energy under grant Contract Number  DE-SC0012567, and in part by the Wade Fund at the School of Science at MIT.  Numerical calculations and some of the analytic derivations  were assisted by GPT-6 Astra, Claude Opus 5.5 and Claude Fable 5.1. 
The authors of this paper are ordered alphabetically.

\appendix

\section{A formal lattice Hamiltonian for a chiral fermion}\label{app:chiral}

Here we formally rewrite the continuum Hamiltonian for a single 1+1d right-moving Majorana--Weyl fermion in terms of fermionic modes associated with sites of a lattice. 
This Hamiltonian is finite-range; in fact, it only has nearest-neighbor couplings, which is to be contrasted with the non-local Hamiltonian in \cite{PhysRevD.26.468,Zakharov:2026nly}. 
There are infinitely many Majorana fermions per lattice site $j$, denoted by $\gamma^{(n)}_j$ with $n\in \mathbb{Z}$. 
However, unlike the non-chiral models presented in the main text, this is not a genuine 1+1d Hamiltonian, and is qualitatively similar to the domain wall fermion. 
For instance, any finite truncation of the flavor index $n$ introduces an extra massless left-moving mode, in accordance with the Nielsen--Ninomiya theorem \cite{Nielsen:1980rz,Nielsen:1981xu,Nielsen:1981hk}.  
 
The Hamiltonian of a right-moving Majorana--Weyl fermion  in the continuum is
\begin{equation}
H_{\text{R}}=-\frac{i}{2}\int dx\,\chi_{\text{R}}\partial_x\chi_{\text{R}}\,,
\qquad \{\chi_{\text{R}}(x),\chi_{\text{R}}(y)\}=\delta(x-y).
\label{eq:HR}
\end{equation}
The lattice fields are obtained from any orthonormal basis of real functions
\(\varphi_{j,n}(x)\) of \(L^2(\mathbb R)\), labeled by a site \(j\) and a flavor index \(n\), through
\begin{equation}
\begin{aligned}
&\gamma_j^{(n)}=\sqrt2\int\varphi_{j,n}(x)\,\chi_{\text{R}}(x)\,\mathrm dx,\\
&\chi_{\text{R}}(x)=\frac1{\sqrt2}\sum_{j,n}\varphi_{j,n}(x)\,\gamma_j^{(n)}.
\label{eq:gammadef}
\end{aligned}
\end{equation}
The lattice fermions satisfy the canonical anticommutation relations $\{\gamma_j^{(n)},\gamma_\ell^{(m)}\}=2\delta_{j\ell}\delta_{nm}$. 
The Hamiltonian written in terms of the lattice fields is then
\begin{equation}
H_{\text{R}}=-\frac{i}{4}\sum_{j,n}\sum_{\ell,m}
\Bigl(\int\varphi_{j,n}\,\partial_x\varphi_{\ell,m}\,d x\Bigr)\gamma_j^{(n)}\gamma_\ell^{(m)} .
\label{eq:HRbasis}
\end{equation}
The coefficient matrix is real and antisymmetric, and it vanishes unless
the supports of \(\varphi_{j,n}\) and \(\varphi_{\ell,m}\) overlap. So the
lattice model is nearest neighbor if each \(\varphi_{j,n}\) is
supported in \([j-1,j+1]\). A convenient choice of such an orthonormal basis is
\begin{equation}
\begin{aligned}
&\varphi_{j,n}(x)=\\
&(-1)^n\sqrt2\,\cos\frac{\pi(x-j)}{2}\,
\cos\Bigl(2\pi\bigl(n+\tfrac14\bigr)\bigl(x-j+\tfrac12\bigr)\Bigr),\\
&\text{for }|x-j|\leq1,
\label{eq:lapped}
\end{aligned}
\end{equation}
and \(\varphi_{j,n}(x)=0\) otherwise. 
The continuum Hamiltonian written in terms of the lattice fields is
\begin{equation}
\begin{aligned}
&H_{\text{R}}=-\frac{i\pi}{4}\sum_j\sum_{n\in\mathbb Z}
\Bigl[n\,\gamma_j^{(n)}\gamma_j^{(-n)}+\bigl(n+\tfrac12\bigr)\gamma_j^{(n)}\gamma_j^{(-n-1)}\Bigr]\\
&
+\frac{i}{2}\sum_j\sum_{n,n'\in\mathbb Z}t_{nn'}\,\gamma_j^{(n)}\gamma_{j+1}^{(n')},
\label{eq:HRlattice}
\end{aligned}
\end{equation}
where
\begin{equation}
t_{nn'}=\frac{n+n'+\frac12}{4(n-n')^2-1}
+\frac{\pi}{2}\Bigl[\bigl(n+\tfrac12\bigr)\delta_{n',-n-1}-n\,\delta_{n',-n}\Bigr].
\label{eq:tlink}
\end{equation}
 
In momentum space, with
\(\gamma_j^{(n)}=L^{-\frac12}\sum_ke^{ikj}\gamma_k^{(n)}\) and
\(\gamma_{-k}^{(n)}=\gamma_k^{(n)\dagger}\), the Hamiltonian is $H_{\text{R}}=\frac14\sum_k\gamma_{-k}^{\mathsf T}h_{\text{R}}(k)\gamma_k$ with
\begin{equation}
\begin{aligned}
&h_{\text{R}}(k)_{nn'}=i\Bigl[-\pi\Bigl(n\,\delta_{n',-n}+\bigl(n+\tfrac12\bigr)\delta_{n',-n-1}\Bigr)\\
&
+t_{nn'}e^{ik}-t_{n'n}e^{-ik}\Bigr].
\label{eq:hRlapped}
\end{aligned}
\end{equation}
With all flavors kept, 
the dispersion relation is exactly linear 
\begin{equation}
E(k)=k+2\pi\mathcal{N},  \quad\mathcal{N}\in\mathbb Z,
\end{equation}
 without the left-moving branch $E(k)=-(k+2\pi\mathcal{N})$. 
 
\begin{figure}[t]
\centering
\includegraphics[width=\columnwidth]{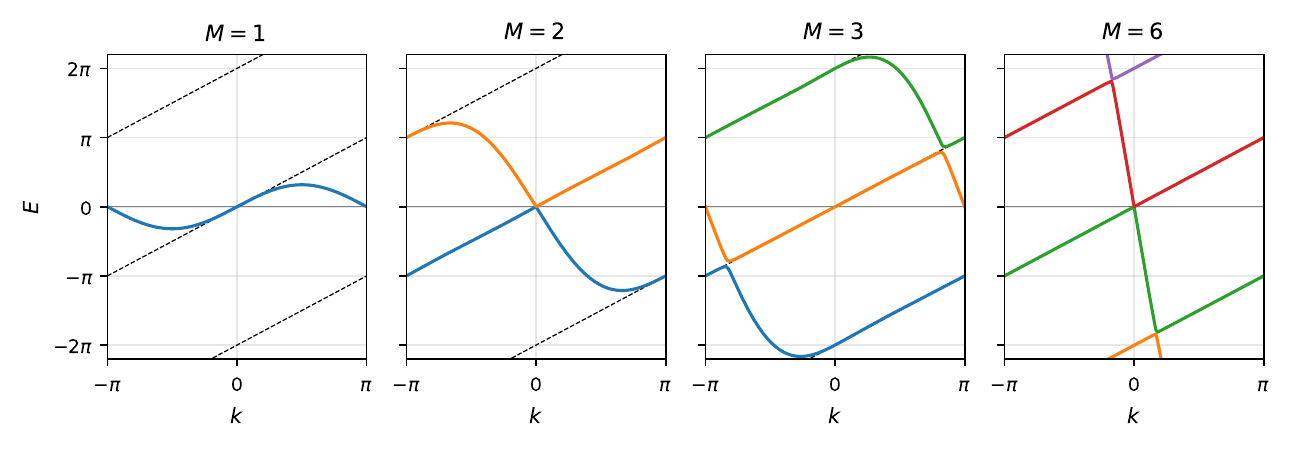}
\caption{Dispersion relation of the truncated 
\(h_{\text{R}}(k)\) in  \eqref{eq:hRlapped} for \(M=1,2,3,6\) Majorana fermions per
site. The continuum linear dispersion relation \(k+2\pi \mathcal{N}\) is shown in dashed.  For odd \(M\) the right mover
through \(k=0\) has unit velocity and the doubler is at \(k=\pi\), with
velocity \(-(2M-1)\); for even \(M\) the doubler is the steep left-moving
band through \(k=0\). The exact model, \(M\to\infty\), has only the dashed branches.}
\label{fig:mw}
\end{figure}

Next, we truncate the lattice Hamiltonian by keeping only $M$ Majorana fermions per site.  
Specifically, we restrict the flavor index to \(n=-p,\ldots,p\)
for \(M=2p+1\) and \(n=-p,\ldots,p-1\) for \(M=2p\). In particular, the $M=1$ truncation gives the critical Kitaev chain \cite{Kitaev:2000nmw}
\begin{equation}
H_{\text{R}}^{(M=1)}=-\frac{i}{4}\sum_j\gamma_j\gamma_{j+1},\qquad
E(k)=\sin k ,
\label{eq:HRM1}
\end{equation}
which has both a left- and a right-moving massless mode. 
In Fig.\ \ref{fig:mw} we show the dispersion relation for $M=1,2,3,6$. In each case we see that finite truncation introduces a left-moving doubler with velocity $-(2M-1)$, similar to the domain wall fermion model.

\begin{figure}
\centering
\includegraphics[width=.4\textwidth]{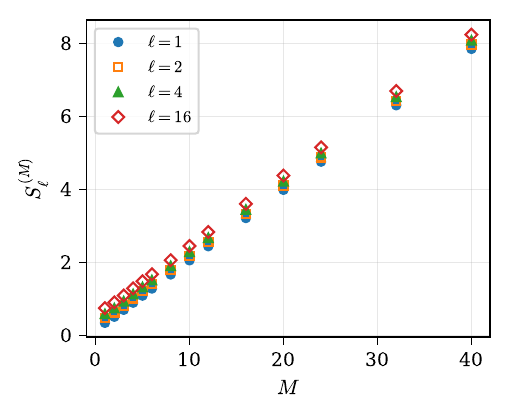}
\caption{Entanglement entropy $S^{(M)}_\ell$ of the first $M$ Majorana modes on an interval of $\ell$ sites in the exact ground state of the chiral lattice model \eqref{eq:HRlattice}, for $L=32$ sites with antiperiodic boundary conditions. The entropy grows linearly in $M$.}\label{fig:chiral_entropy}
\end{figure}

Parallel to Sec.\ \ref{sec:entanglement}, we also compute the entanglement entropy $S^{(M)}_\ell$ of the first $M$ Majorana modes $\gamma^{(n)}_j$ on an interval of $\ell$ sites in the exact ground state, i.e., in the vacuum of the chiral fermion.
For a Gaussian state of Majorana fermions, $S^{(M)}_\ell=-\operatorname{Tr}\bigl[C^{(M)}_\ell\log C^{(M)}_\ell\bigr]$ \cite{peschel2003calculation},  with 
\begin{equation}
C^{(M)}_\ell=\frac12\Bigl(\langle\gamma^{(n)}_j\gamma^{(m)}_{j'}\rangle\Bigr)_{1\le j,j'\le\ell}\,,
\end{equation}
where $n,m$ run over the $M$ kept flavors, and the two-point function in the vacuum on a circle of length $L$ with antiperiodic boundary conditions follows from \eqref{eq:gammadef} and $\langle\chi_\text{R}(x)\chi_\text{R}(y)\rangle=\frac1L\sum_{k>0}e^{ik(x-y)}$:
\begin{equation}
\begin{aligned}
\langle\gamma^{(n)}_j\gamma^{(m)}_{j'}\rangle&=\frac2L\sum_{k>0}e^{ik(j-j')}\,\hat\varphi_n(k)\,\overline{\hat\varphi_m(k)}\,,\\
\hat\varphi_n(k)&=\frac1{\sqrt2}\sum_{\sigma=\pm}e^{i\sigma\pi/4}\Bigl[\frac{\sin\bigl(k+2\pi\sigma(n+\frac12)\bigr)}{k+2\pi\sigma(n+\frac12)}\\
&\qquad\qquad\qquad+\frac{\sin(k+2\pi\sigma n)}{k+2\pi\sigma n}\Bigr].
\end{aligned}
\end{equation}
Here $\hat\varphi_n(k)=\int dx\,\varphi_{0,n}(x)e^{ikx}$ is the Fourier transform of the basis function \eqref{eq:lapped}, and the sum runs over all positive $k\in\frac{2\pi}{L}(\mathbb{Z}+\frac12)$.
As shown in Fig.\ \ref{fig:chiral_entropy}, the entropy does not converge as $M\to\infty$: it grows linearly in $M$, with a slope independent of $\ell$ and $L$, so the entanglement entropy of the exact model is infinite for any interval. 
Moreover, at fixed $M$ the entropy grows linearly in $\ell$  rather than logarithmically. 
These scalings are qualitatively similar to the area law scaling of the entropy for a 2+1d domain wall fermion in a region of $M\times \ell$ sites. 
This is in sharp contrast with the non-chiral models of the main text, where the same quantity converges in $M$ and obeys \eqref{eq:CFTintentanglement}.

\section{The QFT/Lattice map for the Majorana fermion}\label{app:majorana}

Here we provide the explicit map between the continuum Majorana fermion fields and the lattice fields.
We define a new set of real fields $\eta_{\text{L},j}(u),\eta_{\text{R},j}(u)$ as in \eqref{eq:fold} with $\psi$ replaced by $\chi$:
\begin{equation}
\begin{pmatrix}\eta_{\text{R},j}(u)\\ \eta_{\text{L},j-1}(1-u)\end{pmatrix}
=\begin{pmatrix}\cos\theta(u)&\sin\theta(u)\\-\sin\theta(u)&\cos\theta(u)\end{pmatrix}
\begin{pmatrix}\chi_\text{R}(j+u)\\ \chi_\text{L}(j-u)\end{pmatrix},
\end{equation}
which satisfy the same boundary conditions as in \eqref{eq:matching}.
The continuum Hamiltonian written in terms of the new fields takes the form $H=H_{\mathrm{site}}+H_{\mathrm{link}}$ where
\begin{equation}
\begin{aligned}
H_{\mathrm{site}}&=\frac{i}{2}\sum_{j=1}^L\int_0^1 du
\bigl(\eta_{\text{L},j}\partial_u\eta_{\text{L},j}-\eta_{\text{R},j}\partial_u\eta_{\text{R},j}\bigr),\\
H_{\mathrm{link}}&=
i\sum_{j=1}^L\int_0^1 du\,
\theta'(u)\,\eta_{\text{R},j}(u)\,\eta_{\text{L},j-1}(1-u)\,.
\end{aligned}
\end{equation}
Their Fourier expansions are
\begin{equation}
\begin{aligned}
&\eta_{\text{R},j} (u) = {1\over \sqrt{2} }
\sum_{r\in \mathbb{Z}}
\left(
e^{ 2\pi i (r-\frac 14)(u-\frac12)}c_j^{(r)} +\text{h.c.}
\right),\\
&\eta_{\text{L},j} (u) = {1\over \sqrt{2} }
\sum_{r\in \mathbb{Z}}
\left(
i e^{- 2\pi i (r-\frac 14)(u-\frac12)}c_j^{(r)} +\text{h.c.}
\right),
\end{aligned}
\end{equation}
which give the lattice Hamiltonian in \eqref{eq:Hlat} with the same $W_n$.
Combining the above two steps, we find the map between the continuum Majorana fermion fields $\chi_\text{L}(x), \chi_\text{R}(x)$ and the lattice fields $c_j^{(r)}$:
\begin{equation}
\begin{aligned}
c_j^{(r)}=\frac1{\sqrt2}\int_0^1 &du\,\Bigl[
\cos\theta(u)
e^{-2\pi i( r-\frac14)(u-\frac12)}\chi_{\text{R}}(j+u)\\
&
+i\cos\theta(u)\,e^{\,2\pi i( r-\frac14)(u-\frac12)}\chi_{\text{R}}(j+2-u)\\
&+\sin\theta(u) e^{-2\pi i( r-\frac14)(u-\frac12)}\chi_{\text{L}}(j-u)\\
&
-i\sin\theta(u)\,e^{\,2\pi i( r-\frac14)(u-\frac12)}\chi_{\text{L}}(j+u)\Bigr],
\end{aligned}
\label{eq:cfromchi}
\end{equation}
and conversely, for \(0\leq u<1\),
\begin{equation}
\begin{aligned}
&\chi_{\text{R}}(j+u)=\frac1{\sqrt2}\times\\
&\sum_{r\in\mathbb Z}
e^{\,2\pi i( r-\frac14)(u-\frac12)}\Bigl(\cos\theta(u)\,c_j^{(r)}-i\sin\theta(u)\,c_{j-1}^{(r)}\Bigr)
+\mathrm{h.c.},\\
&\chi_{\text{L}}(j-u)=\frac1{\sqrt2}\times\\
&\sum_{r\in\mathbb Z}
e^{\,2\pi i( r-\frac14)(u-\frac12)}\Bigl(\sin\theta(u)\,c_j^{(r)}+i\cos\theta(u)\,c_{j-1}^{(r)}\Bigr)+\mathrm{h.c.}\,.
\end{aligned}
\label{eq:chifromc}
\end{equation}

\bibliographystyle{ytphys}
\bibliography{ref}

\end{document}